\documentclass[prl,twocolumn]{revtex4}
\usepackage{amssymb}
\usepackage{epsfig}

\usepackage{amsmath}
\usepackage{graphicx}

\begin{document}
\title{Unusually strong space-charge-limited current in thin wires}
\author{A. Alec Talin$^{1\ast }$, Fran\c{c}ois L\'{e}onard$^{1}$, B. S.
Swartzentruber$^{2}$, Xin Wang$^{3}$, Stephen D. Hersee$^{3}$}
\affiliation{$^{1}$Sandia National Laboratories, Livermore, California 94551}
\affiliation{$^{2}$Sandia National Laboratories, Albuquerque, New Mexico 87185}
\affiliation{$^{3}$Center for High Technology Materials and Department of Electrical and
Computer Engineering, University of New Mexico,
Albuquerque, NM 87106}
\date{\today }

\begin{abstract}
The current-voltage characteristics of thin wires are often observed to be
nonlinear, and this behavior has been ascribed to Schottky barriers at the
contacts. We present electronic transport measurements on GaN nanorods and
demonstrate that the nonlinear behavior originates instead from
space-charge-limited current. A theory of space-charge-limited current in
thin wires corroborates the experiments, and shows that poor screening in
high aspect ratio materials leads to a dramatic enhancement of space-charge
limited current, resulting in new scaling in terms of the aspect ratio.
\end{abstract}
\maketitle

Nanorods and nanowires made of different semiconducting materials have been
extensively characterized electrically\cite%
{duan,lieber,ham,zhang,reed,dai,yang,huang,fasth}. For most applications, a
linear current-voltage relationship is desirable, usually requiring ohmic
contact to doped wires. However, even in situations where these conditions
should be met, it is often observed that the current-voltage characteristics
are nonlinear. Invariably, this behavior is explained by the presence of
Schottky barriers at the contacts, despite the fact that such models
sometimes give poor descriptions of the experimental data. Properly
identifying the factors that influence electrical transport characteristics
is important for device design but also because extraction of material
parameters such as the mobility relies on analysis with specific models.

It is well-known that in bulk insulating and semiconductor materials,
space-charge-limited (SCL) current leads to nonlinear, non-exponential $I-V$
characteristics\cite{grinberg,lampert,smith} with the relationship $%
I\propto V^{2}$. This behavior occurs in situations of mobility-dominated
transport when the effective carrier concentration is low. This can arise
due to low intrinsic doping, charge traps, or depletion widths at the
contacts that are larger than the channel length (punchthrough). Thin wires
should be particularly sensitive to SCL effects for several reasons: first,
because electrostatic screening in high aspect ratio systems is poor\cite%
{leonard}, the injected carriers cannot be effectively screened; second,
carrier depletion due to surface states is expected to be more important in
thin wires due to the large surface-to-volume ratio\cite{gu}; third, charge
traps may readily be incorporated during growth\cite{schricker}.

In this letter, we present electrical transport in individual GaN nanorods
showing symmetric, nonlinear $I-V$ characteristics. We show that the
relationship $I\propto V^{2}$ is satisfied in these thin wires, a
signature of SCL current. A theory for SCL transport in thin wires is
presented and shows that SCL current is unusually strong due to a new
scaling with the wire aspect ratio.

The growth and microstructural characterization of the GaN nanorods has been
described in detail elsewhere\cite{hersee}. Briefly, the nanorods were grown
in a commercial metal-organic chemical vapor-deposition system on
GaN/sapphire substrates using selective epitaxy, whereby a 30 nm thick SiN
film with lithographically defined holes serves as a mask for the nanorod
growth. For this work, the nanorod aspect ratio $R/L$ ranged from 0.05 to
0.5, and the nanorods were up to a few microns in length. Electrical
measurements on individual nanorods were carried out in two ways. In the
first approach, electrical contacts were defined on top of randomly
dispersed nanorods on SiO$_{2}$ substrates using optical lithography
followed by electron beam evaporation of Ti/Au (10nm/300nm) and lift-off.
The electrode pattern consists of arrays of interdigitated, individually
addressable electrodes with spacing ranging from 1 $\mu $m to 4 $\mu $m. In
the second approach, the vertical nanorods were individually contacted at
their top end directly on the growth wafer using a Au-coated W STM tip, with
a large area Ag paint serving as the back electrode.

Figure 1a shows a scanning electron microscope (SEM) image of an individual
GaN nanorod with two Ti/Au side contacts, as well as the associated $I-V$
characteristics for several of these devices. The $I-V$ curves are clearly
nonlinear and are fairly symmetric between positive and negative values of
the source-drain voltage. An excellent description of the data is obtained
by plotting $I/V$ vs $V$ as shown in Fig. 1b. There, several of the positive
and negative $I-V$ traces of Fig. 1a are plotted, with the current
normalized by that measured at 3 Volts. Clearly, straight lines are
obtained, indicative of a $I\propto V^{2}$ behavior; the values of the
intercept suggest that a linear contribution to the $I-V$ curve is small.
The quadratic $I-V$ characteristics cannot be explained by Schottky or
tunneling barriers at the contacts. Indeed, Ti is often used to make low
resistance contacts to bulk GaN\cite{ruterana} and has been shown to give
ohmic contacts to GaN nanowires even in the absence of annealing\cite{reed}.
In addition, we also annealed several of the devices at 550 $%
{{}^\circ}%
$C in vaccum for 15 minutes, and found essentially no change in the $I-V$
characteristics. Furthermore, for back-to-back Schottky barriers, the
current is dominated by the reverse-biased Schottky barrier, and has the
form $1-\exp \left( -eV/kT\right) $ in the absence of tunneling; this gives
a linear $I-V$ curve at low bias, and saturation at larger bias, failing to
even qualitatively describe our data. If tunneling is allowed, then explicit
calculations for back-to-back Schottky contacts give an exponential behavior %
\cite{zhang}. This is also born out in the well-known expressions for the $%
I-V$ characteristics of two tunnel contacts which lead to a linear behavior
at low bias and an exponential behavior at high bias\cite{sze}.

\begin{figure}[h]
\centering
\includegraphics[width=8cm]{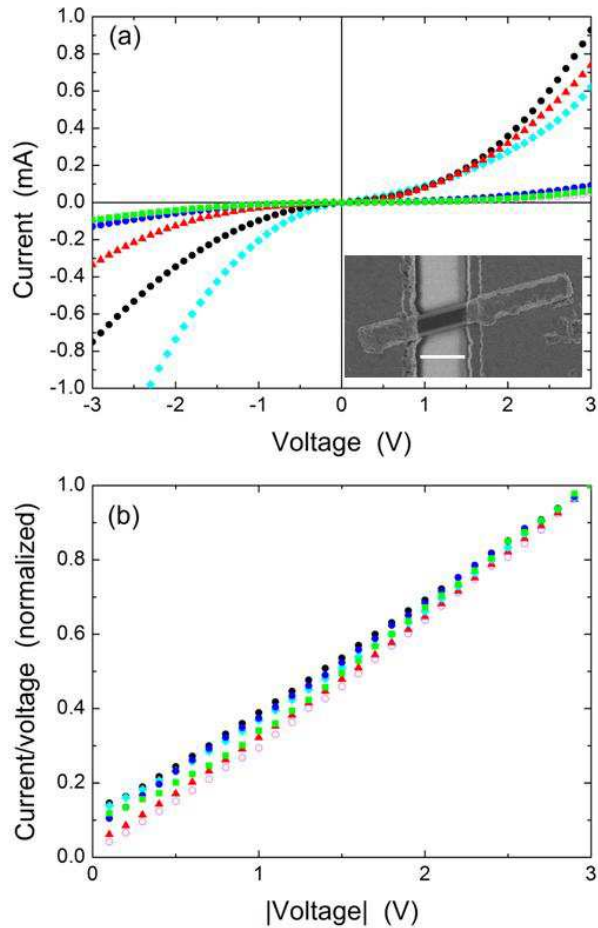}
\caption{(Color online) (a) Current-voltage characteristics of several GaN
nanowires. Inset shows a SEM image of one of the devices. Scale bar is one
micron. (b) Data of panel (a) but plotted as current divided by voltage,
versus voltage. The data for several positive and negative voltage sweeps
from several different devices is plotted, and normalized to 1 for an
absolute voltage of 3 Volts.}
\end{figure}

To further support the argument that contacts are not responsible for the
observed behavior, we used a Au-coated W STM tip to directly contact
individual nanorods on the growth substrate, and probe the transition from
injection-limited to SCL behavior. As we indicated earlier, a large area,
low resistance Ag contact is painted on the back of the growth substrate;
thus, the current is dominated by the nanorod itself, or by the small area
nanorod/tip contact. Figure 2 shows the $I-V$ curves and a SEM image for a
particular nanorod. When the tip is initially brought into contact with the
nanorod, the $I-V$ characteristics are rectifying and well described by
transport through a barrier, i.e. the $I-V$ curve is exponential at forward
bias, see bottom inset of Fig. 2. This corresponds to the injection-limited
regime, and the $I-V$ is dominated by the behavior of the nanorod/tip
contact. As the tip is pressed into the nanorod, the nature of the $I-V$
characteristics changes, with the current increasing by a factor of 100,
becoming symmetrical about the origin, and well described by a $I\propto
V^{2}$ behavior. These measurements demonstrate unequivocally the transition
between injection-limited and SCL transport; since there is only one
relevant contact in these measurements, the symmetrical current-voltage
characteristics must be due to the nanorod itself. (The reason for the
improvement in the tip/nanorod contact quality is not entirely clear, but
could be due to a combination of a thin native oxide being destroyed as the
tip is pressed into the nanorod and the small radius of curvature of the tip
leading to substantial field enhancement. It should be noted that in several
nanorods the $I-V$ characteristics changed gradually from rectifying to
leaky diode and finally to SCL.)

\begin{figure}[h]
\includegraphics[width=8cm]{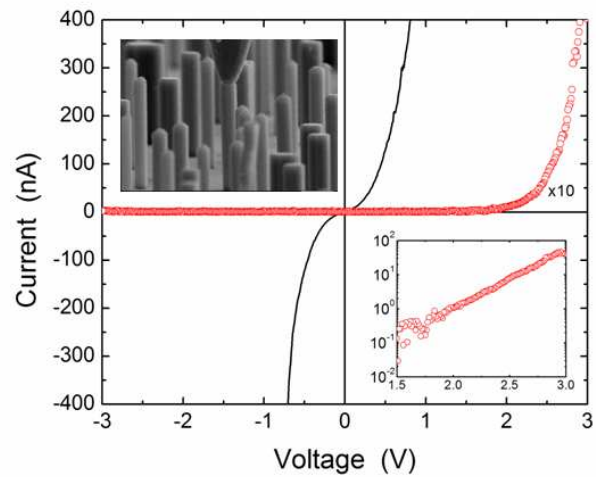}
\caption{(Color online) Current-voltage characteristics before (symbols) and
after (line) the tip is pressed into the nanowire. The bottom right inset
shows the current before pressing on a log scale. The top left inset is a
SEM image of the Au-coated W tip contacting a GaN nanowire.}
\end{figure}

The observations of quadratic $I-V$ curves are consistent with a theory of
SCL current in thin wires. For transport through a thin wire of radius $R$
and length $L$, the drift component of the current density is%
\begin{equation}
J=\frac{2}{R^{2}}\int_{0}^{R}rj\left( r\right) dr=\frac{2e\mu }{R^{2}}%
\int_{0}^{R}rn(z,r)E_{z}(z,r)dr  \label{current}
\end{equation}%
where $j\left( r\right) =$ $e\mu n(z,r)E_{z}(z,r)$ is the current density at
radial coordinate $r$, $\mu $ is the mobility, $n(z,r)$ the carrier
concentration, and $E_{z}(z,r)$ the electric field in the transport
direction $z$. (The diffusion component of the current is negligible for
voltages $\gtrsim 10kT/e$\cite{grinberg}) This equation is combined with
Poisson's equation for the electrostatic potential {\it inside} the thin wire

\begin{equation}
\frac{\partial ^{2}V}{\partial r^{2}}+\frac{1}{r}\frac{\partial V}{\partial r%
}+\frac{\partial ^{2}V}{\partial z^{2}}=\frac{en\left( z,r\right) }{\epsilon 
}  \label{poisson}
\end{equation}%
where $\epsilon $ is the semiconductor dielectric constant, and where we
assumed that the charge distribution is isotropic in the angular coordinate $%
\phi $. (Note that outside the wire $r>R$, we have $n(z,r)=0$.) Solution of
Poisson's equation with the Green's function $\left( 4\pi \right)
^{-1}\left| {\bf r}-{\bf r}^{\prime }\right| ^{-1}$ \cite{GF} gives the
spatially-dependent electric field {\it inside} the wire%
\begin{equation}
E_{z}(z,r)=\frac{e}{\epsilon }\int_{0}^{L}\int_{0}^{R}n(z^{\prime
},r^{\prime })G(z-z^{\prime };r,r^{\prime })r^{\prime }dr^{\prime
}dz^{\prime },
\end{equation}%
where $G(z-z^{\prime };r,r^{\prime })=\left( 4\pi \right) ^{-1}\partial
_{z}\int_{0}^{2\pi }\left( \left| {\bf r}- {\bf r}^{\prime }\right|
^{-1} - \right.$ $\left. \left| {\bf r}-{\bf L}\right| ^{-1} \right) d\phi ^{\prime }$ 
with ${\bf L\ }$a point on the electrode at $z=L$. When combined with Eq. $\left( %
\ref{current}\right) $, this expression for the electric field provides an
integral equation\cite{grinberg} for the spatially-dependent carrier
concentration; with the scalings $\xi =z/L$, $\alpha =r/L$ and $\nu \left(
\xi \right) =\left( 2\epsilon J/e^{2}\mu L\right) ^{-1/2}n\left( \xi \right) 
$ this equation is $\left( 4L^{2}/R^{2}\right) \int_{0}^{R/L}\nu (\xi
,\alpha )\alpha d\alpha \left[ \int_{0}^{R/L}\int_{0}^{1}\nu (\xi ^{\prime
},\alpha ^{\prime })G\left( \xi \right. \right.$ $\left. \left. -\xi ^{\prime },\alpha ,\alpha ^{\prime
}\right) \alpha ^{\prime }d\alpha ^{\prime }d\xi ^{\prime }\right] =1$. The
potential on the drain electrode can be obtained from $V=-%
\int_{0}^{L}E_{z}(z,r)dz$ and is given by
\vspace{-5pt}
\begin{equation}
V=-\left( \frac{JL^{3}}{2\epsilon \mu }\right) ^{1/2}\int_{0}^{1}\left[ 
\frac{L^{2}}{2R^{2}}\int_{0}^{R/L}\alpha d\alpha \nu (\alpha ,\xi )\right]
^{-1}d\xi .
\end{equation}%
The integral in this last equation is simply a numerical factor, which we
define as $\left( \zeta /2\right) ^{-1/2}$; the total current density is then%
\begin{equation}
J=\zeta \left( R/L\right) \frac{\epsilon \mu }{L^{3}}V^{2}.  \label{density}
\end{equation}%
Thus, we find that for a thin wire, the current density depends
quadratically on the applied voltage, with a prefactor that depends on the
ratio $R/L$. This can be compared with SCL current in a bulk material by
using $n(z,r)=n(z)$ and taking the limit $R/L\gg 1$ in the above equations,
giving $\zeta =9/8$ and the current density \cite{grinberg,lampert} $%
J_{bulk}=\left( 9\epsilon \mu /8L^{3}\right) V^{2}$. The important point is
that the dimensionality does not change the dependence on voltage, but
affects the scaling with length. In any dimension where SCL current
dominates, a plot of $I/V$ vs $V$ gives a straight line, as observed in our
experimental data. (Velocity saturation would lead to $I\propto V$ at
larger voltages, which we have not seen in our measurements; this is
consistent with electric fields larger than 10 V/$\mu m$ needed to reach the
maximum carrier velocity in GaN\cite{pearton}.)

Additional evidence for SCL transport comes from a more detailed study of
the experimental data in terms of a scaling analysis. Equation $\left( \ref%
{density}\right) $ can be re-written as%
\begin{equation}
\frac{L^{3}}{AV^{2}}I=\epsilon \mu \zeta \left( R/L\right)
\end{equation}%
where $A$ is the nanowire cross-sectional area, and $\zeta $ is the scaling
function. Figure 3 shows a plot of $(IL^{3})/(AV^{2})$ as a function of $R/L$
for a number of nanorods contacted lithographically and with the STM tip. It
is remarkable that both types of contacts, despite their very different
geometry, lead to the same universal behavior, providing further evidence
that the observed behavior is not due to contacts. Furthermore, the
dependence of the data on $R/L$ is entirely consistent with the theoretical
analysis presented above. Indeed, for small $R/L$ the carrier concentration
is independent of $r$ to a good approximation\cite{radial}, and we can
derive that $\zeta \left( R/L\right) =\zeta _{0}\left( L/R\right) ^{-2}$; a
similar situation holds for large $R/L$ and we have $\zeta \left( R/L\right)
\sim const$. The solid lines in the main figure and the inset show that a
fit of the form $a\left( L/R\right) ^{-2}$ represents well the data for $%
R/L<0.2$; the crossover to the large $R/L$ regime is also apparent. This
explicit dependence of the current on the length of the nanorod $L$ also
indicates that the behavior is not due to contacts. In addition, from the
relation $a=\zeta _{0}\epsilon \mu $ and the fit to the experimental data,
the effective carrier mobility is calculated to be~386 cm$^{2}$/Vs\cite{zeta}%
. This value is consistent with values for bulk GaN material at room
temperature\cite{pankove}, again supporting the conclusion that the bulk of
the nanorod itself determines the behavior.

\begin{figure}[h]
\includegraphics[width=8cm]{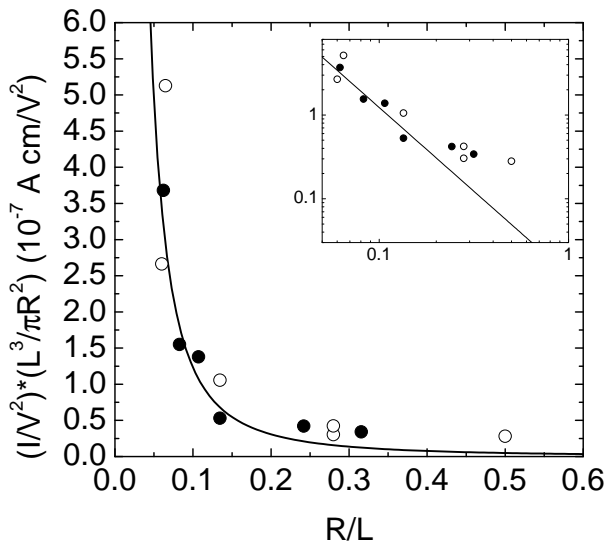}
\caption{Scaling behavior for several GaN nanowires with
lithographically-defined top contacts (solid circles) and with the STM tip
as the contact (open circles). The inset is the same data plotted on a
log-log scale. In both plots the solid line is a best fit to the data of the
form $a(R/L)^{-2}$ for $R/L<0.2$.}
\end{figure}

In bulk materials, there exists a cross-over voltage $V_{c}^{bulk}$ between
an initial ohmic behavior and SCL transport. For small voltages, the current
has an ohmic behavior $J=e\mu n_{0}V/L$ where $n_{0}$ is the effective
carrier concentration. The transition to the SCL regime occurs when $%
J_{bulk}^{SCL}=\left( 9\epsilon \mu /8L^{3}\right) V^{2}$ exceeds the ohmic
current, or $V_{c}^{bulk}\sim en_{0}L^{2}/\epsilon $. A similar argument
holds for thin wires, except that the cross-over voltage is given by%
\begin{equation}
V_{c}=\frac{1}{\zeta \left( R/L\right) }\frac{en_{0}L^{2}}{\epsilon }\sim 
\frac{1}{\zeta \left( R/L\right) }V_{c}^{bulk}\text{.}
\end{equation}%
The presence of the factor $\zeta ^{-1}\left( R/L\right) $ has a significant
impact on the magnitude of $V_{c}$. Indeed, for small $R/L$ we have $%
V_{c}\sim \left( R/L\right) ^{2}V_{c}^{bulk}$ and the crossover voltage is
reduced by orders of magnitude for small values of $R/L$. The immediate
implication is that the carrier concentration needed to obtain Ohmic
behavior is much larger in thin wires, by a factor $(L/R)^{2}$. And more
generally, it points to the conclusion that, as long as carrier injection is
efficient, SCL transport should be prevalent in materials at reduced
dimensionality. The origin of this behavior lies in the fact that in
high-aspect ratio structures, the Coulomb interaction is poorly screened\cite%
{leonard,achoyan}. For SCL transport in thin wires, this means that the
effective magnitude of the injected charge is significantly increased
because it is only screened up to a radius $R$.

In summary, we find that GaN nanorods show nonlinear $I-V$ characteristics
of the form $I\propto V^{2}$ which we ascribe to SCL transport. This
electronic transport regime arises when conduction is mobility-dominated,
when the carrier injection is efficient, and the carrier concentration is
low. This situation may be prevalent in thin wire materials for several
reasons. First, because of the poor electrostatic screening in thin wires,
SCL current is enhancend by orders of magnitude. Second, dopant
incorporation during growth is often difficult to achieve, so doping levels
may be inherently low, leading to long depletion widths and punchthrough, a
situation that is exacerbated by the poor electrostatic screening. In
addition, surface states and charge traps may deplete the carriers. Finally,
because of the scaling of the SCL current with the aspect ratio, much larger
effective carrier concentrations are needed to obtain ohmic behavior. Taken
together, these arguments suggest that future electronic devices based on
nanorods and nanowires should carefully consider SCL behavior in their
design.

Sandia is a multiprogram laboratory operated by Sandia Corporation, a
Lockheed Martin Company, for the United States Department of Energy under
Contract No. DEAC01-94-AL85000. Work performed in part at the U.S.
Department of Energy, Center for Integrated Nanotechnologies, at Los Alamos
National Laboratory (Contract No. DE-AC52-06NA25396) and Sandia National
Laboratories (Contract No. DE-AC04-94AL85000).

$^{\ast }$email:aatalin@sandia.gov


\begin{references}
\bibitem{GF} The Green's function depends on the boundary conditions on the
electrodes, and the expression used here is appropriate for point contacts,
with the compensating charge on the drain. Calculations for more complicated
electrode geometries do not change the $I\propto V^{2}$ behavior, only the
numerical prefactor. This is discussed in detail for thin films in Ref. \cite%
{grinberg}.
\bibitem{zeta} To obtain this value of the mobility, we used $\zeta _{0}=1$.
This is consistent with calculations in different thin film and bulk
geometries \cite{grinberg}, where $\zeta _{0}$ is always close to 1.
\end{references}
\end{document}